# Half-Duplex Attack: An Effectual Attack Modelling in D2D Communication

Misbah Shafi
Student Member IEEE
SMVDU, J&K, India
misbahshafi0@gmail.com

Rakesh Kumar Jha
Senior Member IEEE
SMVDU, J&K, India
jharakesh.45@gmail.com

*Abstract*—The visualization of future generation Wireless Communication Network (WCN) redirects the presumption of onward innovations, the fulfillment of user demands in the form of high data rates, energy efficiency, low latency, and long-range services. To content these demands, various technologies such as massive MIMO (Multiple Input Multiple Output), UDN (Ultra Dense Network), spectrum sharing, D2D (Device to Device) communication were improvised in the next generation WCN. In comparison to previous technologies, these technologies exhibit flat architecture, the involvement of clouds in the network, centralized architecture incorporating small cells which creates vulnerable breaches initiating menaces to the security of the network. The half-duplex attack is another threat to the WCN, where the resource spoofing mechanism is attained in the downlink phase of D2D communication. Instead of triggering an attack on both uplink and downlink, solely downlink is targeted by the attacker. This scheme allows the reduced failed attempt rate of the attacker as compared to the conventional attacks. The analysis is determined on the basis of Poisson's distribution to determine the probability of failed attempts of half duplex attack in contrast to a full duplex attack.

*Index Terms*—half duplex attack, failed attempt rate, D2D, WCN, security, Poisson's process.

## I. INTRODUCTION

Research on security occupies the vital aspect of WCN. Various security issues were revealed from preceding years though security attacks on WCN is an approaching advent that requires immense attention. Based on the fundamentals of security, security attacks can be characterized as attacks on identity protection, location privacy, communication confidentiality, availability, authorization, and authentication [1]. Some of the attacks that are closely associated with 5G (Fifth Generation) WCN include bandwidth spoofing, IP (Internet Protocol) spoofing, pilot contamination attack, rouge access point attack, signaling storms, ping flood, radius cracking [2]. D2D communication defines the shortest communication between device nodes without any requirement to pass across the core network or access point [3]. The operations involved in D2D communication are self-managing and spontaneous. Therefore, complete security adaptability is still its challenging aspect in comparison to the security provided by the centralized cellular network [4].

## II. STATE OF THE ART

D2D allows the devices to perform communication with each other independently, devoid of network infrastructure. It leverages several vulnerable security threats such as physical attacks, compromised credentials, configuration attacks, protocol attacks, and privacy attacks [4]. Possible security attacks in D2D communication include proximity based attack [5], free-riding attack, inference attack, trust forging attack, impersonation, location spoofing, data fabrication attack, and DDoS (Distributed Denial of Service) [6], fake profile attacks [7]. These attacks target both downlink and uplink to execute the operation. To deteriorate the rate of intrusion detection, one of the probable is the execution of the Half-Duplex (HD) attack.

### A. Contribution

This paper proposes an attack based on the downlink solely. The energy efficiency, probability of failed attempts, and failed attempt rate of the HD attack are compared with the full duplex attack

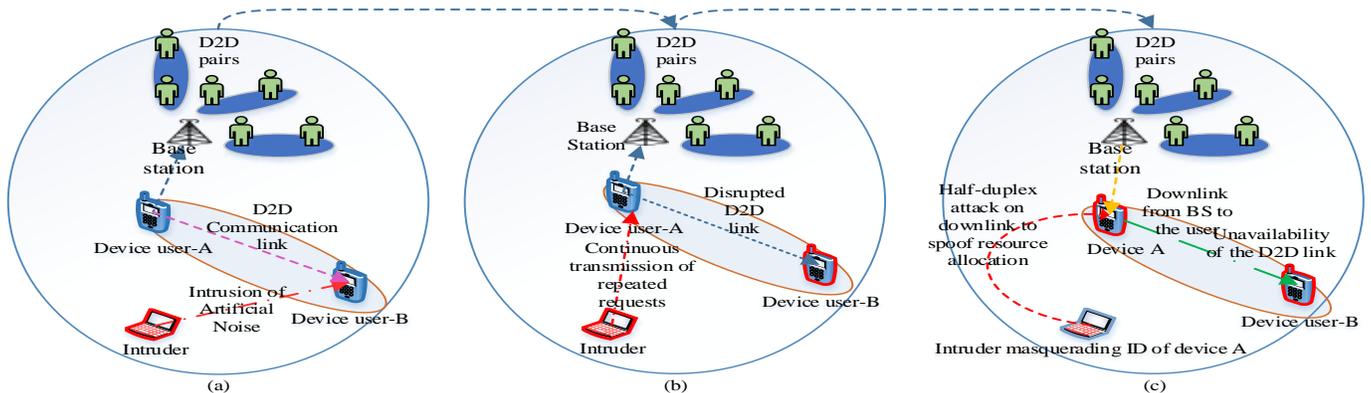

Fig. 1. (a) Phase of AN forwarding (b) Phase of continuous requests (c) Phase of resource spoofing

## III. Research Proposal

This proposal integrates Half Duplex (HD) attack on the relay device of the D2D pair. These devices are comparatively less secure as compared to other communication identities. HD attack scenario involves the D2D communication with device relaying controlled link assisted by BS (Base Station). It encompasses the triggering of an attack on downlink specifically. It is observed that HD attack occupies more probability to occur with more sensitivity and less complexity as compared to conventional Full Duplex (FD) attacks.

### A. System model

Consider a device node $B$ present the edge of the cell where the coverage of the reception is comparatively weak. Following the advantage of D2D backed communication, another device node $A$ present in the reachable zone of signal from the BS. Thereby, device $A$ acts as a relay, and BS is liable to allocate resources to the devices. Assume the D2D communication takes place in the presence of an intruder with an unknown CSI (Channel State Information). HD attack comprises of three phases, and each phase is diagrammatically shown in Fig. 1. These phases can be defined as:

#### 1) The phase of AN (Artificial Noise) forwarding

It involves the intrusion of AN by the intruder $I$ to the device node $B$ present at the boundary of the cell. The purpose of incurring AN to device $B$ is to create noise during the D2D communication phase such that artificially generated noise is transmitted along with the information signal, as shown in Fig. 1(a). Thereby creating the interference effect on the device $B$.

#### 2) The phase of continuous requests

The second stage involves the continuous transmission of requests for the availability of resources to form a D2D pair at the device $A$ by the intruder $I$, as shown in Fig. 1.(b). The main objective of this phase is to drain the resources of the device $A$ and make it unavailable for the downlink phase of resource allocation. The mechanism is incorporated after the transmission of authentication request by the device $A$ to the BS. The continuous procedure of repeated requests to the device $A$ results in the failure of its availability. It is followed by the response of the device $A$ to the intruder.

#### 3) The phase of resource spoofing

It involves the response attained by an intruder from the BS intended for the valid device $A$ as shown in Fig. 1(c). The phase involves the masquerading the identification of valid device $A$ by the intruder $I$ such that intruder receives the information present in the downlink from BS to the intruder. Therefore, takes hold of the allocated resources, which were meant to be allocated to the legitimate device $A$. The HD attack is evaluated by the parameter of failed attempt probability. Let $x_1^n$ as the total number of attempts made by the intruder to attack the device-A, $\phi$ as the mean of $x_1^n$, $w_{dl}^{HD}$ and $w_r^{FD}$ as the number of failed attempts to access resource allocation in HD attack and FD attack such that the Failed Attempt Rate (FAR) for the HD attack, $W_r^{HD}$ can be expressed as:

$$W_r^{HD} = \frac{w_{dl}^{HD}}{x_1^n} \quad (1)$$

For FD the FAR $W_r^{FD}$ can be obtained as:

$$W_r^{FD} = \frac{w_r^{FD}}{x_1^n} \quad (2)$$

The above equation can also be re-written as:

$$W_r^{FD} = \frac{w_{ul}^{FD} + w_{dl}^{FD}}{x_1^n} \quad (3)$$

For the HD attack, the number of failed attempts to receive the downlink successfully are defined for downlink only. Let the number of the failed attempts in the downlink in HD attack be equal to the number of failed attempts in the downlink in FD attack such that equation (3) can be given as:

$$W_r^{FD} = \frac{w_{ul}^{FD} + w_{dl}^{HD}}{x_1^n} \quad (4)$$

$$= \frac{w_{ul}^{FD}}{x_1^n} + W_r^{HD} \quad (5)$$

Where $w_{ul}^{FD}, w_{dl}^{FD}$ are the number of attempts in uplink and downlink respectively for the FD attack, $\frac{w_{ul}^{FD}}{x_1^n}$ is a non-negative quantity whose value lies between 0 and 1. Therefore, form equation (5) it can be concluded as:

$$W_r^{FD} \geq W_r^{HD} \qquad \forall \frac{w_{ul}^{FD}}{x_1^n} \geq 0 \quad (6)$$

Equation (6) justifies the analysis that the FAR achieved in the FD attack is more than the FAR obtained in HD attack. Then the probability of the random variable characterized by the FAR in HD, using Poissons distribution is given by:

$$P(w_{dl}^{HD}; \phi) = \frac{e^{-\phi}\phi^{w_{dl}^{HD}}}{w_{dl}^{HD}!} \quad (7)$$

Probability of failed attempts for uplink in full duplex attack is expressed as:

$$P(w_{ul}^{FD}; \phi) = \frac{e^{-\phi}\phi^{w_{ul}^{FD}}}{w_{ul}^{FD}!} \quad (8)$$

Probability of failed attempts for downlink in full duplex attack is assumed to be equal to the probability of failed attempts for downlink in HD attack, such that the probability of failed attempts in FD attack can be expressed as:

$$P(w_r^{FD}; \phi) = P(w_{ul}^{FD}; \phi) + P(w_{dl}^{HD}; \phi) - P(w_{ul}^{FD}; \phi) \cap P(w_{dl}^{HD}; \phi) \quad (9)$$

The occurrence of each attempt is independent at each time interval such that $P(w_{ul}^{FD}; \phi)$ and $P(w_{dl}^{HD}; \phi)$ are mutually exclusive.

Therefore, $\quad P(w_{ul}^{FD}; \phi) \cap P(w_{dl}^{HD}; \phi) = 0 \quad (10)$

Using equation (10) in equation (9), we get

$$P(w_r^{FD}; \phi) = P(w_{ul}^{FD}; \phi) + P(w_{dl}^{HD}; \phi) \quad (11)$$

The probability of failed attempts in full duplex attack is less than the probability of failed attempts in HD attack given by:

$$P(w_r^{FD}; \phi) \geq P(w_{dl}^{HD}; \phi) \quad \forall P(w_{ul}^{FD}; \phi) \geq 0 \quad (12)$$

The equation (12) identifies that the failed attempt probability of the FD attack under the impact of AD is always greater than the probability of the failed attempts in HD attack.

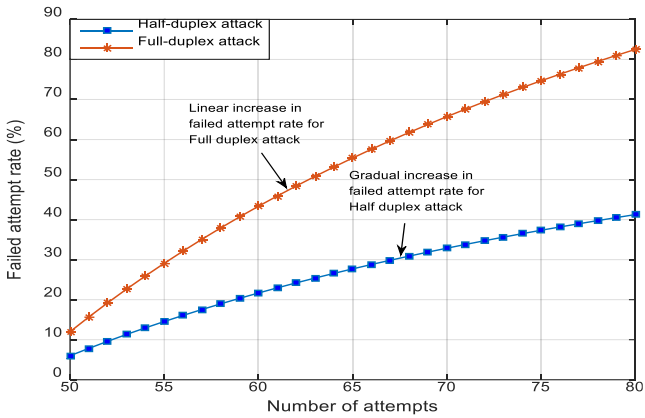
Fig. 2. Failed attempt rate versus number of attempts

## IV. SIMULATION RESULTS

In this section, simulation results are elaborated in the MATLAB simulation environment, and the conclusions attained are being stated for half-duplex attack and full-duplex attack. The simulation parameter involves the bandwidth of 800MHz and the user distance of 150m to 250m from the base station. Fig. 2. demonstrates the results of FAR concerning the number of attempts for the half-duplex and full-duplex attack. From the graphical examination of Fig. 2, it is analyzed that FAR shows an approximate linear behavior with respect to the number of attempts taken to attack the device. Moreover, high FAR for FD attack is observed as compared to the HD attack for an increasing number of attempts. In Fig. 3, it is detected that the probability of occurrence of failed attempts is more in FD attack than in HD attack. It is also perceived that the probability of failed attempts increases with an increase in a number of failed attempts initially and from the maximum probability, it decreases with an increase in the failed attempts representing the bell-shaped curve characteristics. The results of probability are obtained by using Poisson's distribution, as mentioned in equation (7) and (8). Fig. 4 represents the energy efficiency analysis of HD and FD attack in the urban and rural scenarios. It is illustrated that in both scenarios of rural and urban communication network HD attack is more energy-efficient than the FD attack. Furthermore, HD attack in rural scenario occupies more energy efficiency than in urban scenario. Therefore, it can be concluded that the HD attack is predicted to be more effective than an FD attack.

## V. CONCLUSION

In this paper, possible attack strength is explored, wherein solely downlink is targeted in the form of an HD attack. The proposed attack is compared with the FD attack involving both uplink and downlink. This approach defines the security perspective of attacking attempts in an HD mode and in FD mode. In HD mode, the attempt of spoofing the resources meant to be allocated to the legitimate user is allocated to the intruder instead. The probability of occurrence of failed attempts in HD attack is reflected to be less and correspondingly strengthening the successful attacking attempts as compared to the FD attack. The phases of the attacking strategy in HD mode are explained

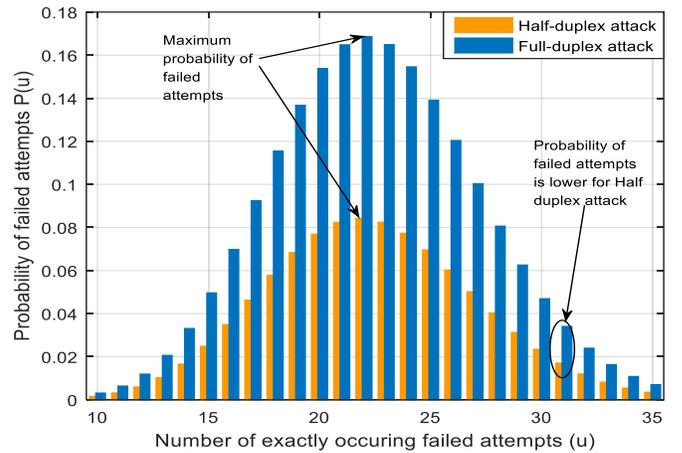
Fig. 3. Probability of failed attempts versus number of exactly occurring attempts

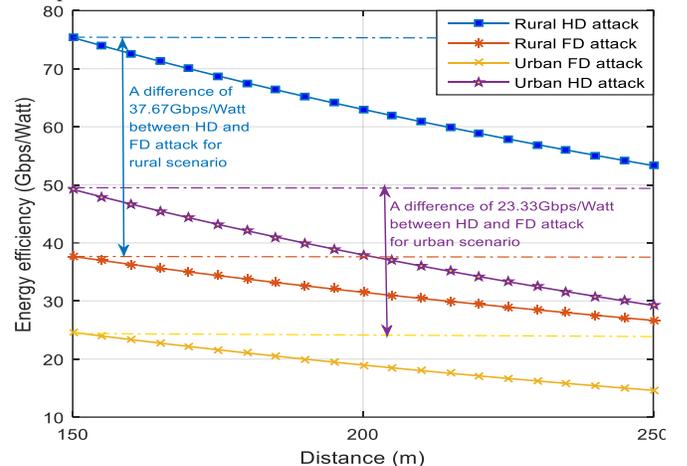
Fig. 4. Energy Efficiency versus Distance for HD (Half-Duplex) attack ad FD (Full Duplex) attack under rural and urban scenarios

schematically. Further, the simulation results are implicated, demonstrating the effectiveness of the HD attack in comparison to the FD attack. It is reflected from the parameter analysis of energy efficiency, FAR, and the probability of false attempts the proposed attack is more effectual than the FD attack.